%% file: main.tex
\documentclass[conference]{IEEEtran}
\IEEEoverridecommandlockouts
\usepackage{cite}
\def\BibTeX{{\rm B\kern-.05em{\sc i\kern-.025em b}\kern-.08em
		T\kern-.1667em\lower.7ex\hbox{E}\kern-.125emX}}

\usepackage[american]{babel}
\usepackage[utf8]{inputenc}
\usepackage[T1]{fontenc}
\usepackage{enumitem}

\usepackage{mathtools}
\usepackage{amssymb}
\usepackage{amsthm}

\DeclareMathOperator*{\argmin}{argmin}

\usepackage[hyphens]{url}

\usepackage{tabu}
\usepackage{soul}
\usepackage{booktabs}

\usepackage{graphicx}
\usepackage{subcaption}

\usepackage{algorithm}
\usepackage{algpseudocode}

\usepackage[variablett]{lmodern}
\usepackage{listings}
\makeatletter
\newbox\sf@box
{\def\sf@one{#1}%
	\def\sf@two{#2}%
	\setbox\sf@box\hbox
	\bgroup}%
{ \egroup
	\ifx\@empty\sf@two\@empty\relax
	\def\sf@two{\@empty}
	\fi
	\ifx\@empty\sf@one\@empty\relax
	\subfloat[\sf@two]{\box\sf@box}%
	\else
	\subfloat[\sf@one][\sf@two]{\box\sf@box}%
	\fi}
\makeatother

\usepackage[xindy, acronym, nomain, toc]{glossaries}
\usepackage{hyperref}

\usepackage[binary-units = true]{siunitx}
\def\BibTeX{{\rm B\kern-.05em{\sc i\kern-.025em b}\kern-.08em
		T\kern-.1667em\lower.7ex\hbox{E}\kern-.125emX}}

\allowdisplaybreaks

\addto\extrasamerican{

}

\begin{document}

\title{\Large ANDREAS: Artificial intelligence traiNing scheDuler foR \\  accElerAted resource clusterS}

\author{
    \IEEEauthorblockN{Federica Filippini, Danilo Ardagna, \\ Marco Lattuada, Edoardo Amaldi}
    \IEEEauthorblockA{
        \textit{Dipartimento di Elettronica, Informazione e Bioingegneria} \\
        \textit{Politecnico di Milano}, Italy \\
        name.lastname@polimi.it}
    \and
    \IEEEauthorblockN{Maciek Riedl, Katarzyna Materka, Paweł Skrzypek}
    \IEEEauthorblockA{
        \textit{7bulls}, Poland \\
        \{mriedl, kmaterka, pskrzypek\}@7bulls.com}
    \and
    \IEEEauthorblockN{Michele Ciavotta}
    \IEEEauthorblockA{
        \textit{Dipartimento di Informatica, Sistemistica e Comunicazione} \\
        \textit{Università degli studi di Milano-Bicocca}, Italy \\
        michele.ciavotta@unimib.it}
    \and
    \IEEEauthorblockN{Fabrizio Magugliani, Marco Cicala}
    \IEEEauthorblockA{
        \textit{E4 Computer Engineering}, Italy \\
        \{fabrizio.magugliani, marco.cicala\}@e4company.com}
}

\maketitle

\begin{abstract}
\input{sections/abstract}
\end{abstract}


\section{Introduction}\label{sec:intro}
\input{sections/intro}

\section{Related work}\label{sec:related}
\input{sections/related}

\section{ANDREAS Architecture}\label{sec:architecture}
\input{sections/architecture}

\section{ANDREAS Optimizer}\label{sec:optimizer}
\input{sections/model}\label{sec:model}
\input{sections/algorithm}

\section{Experimental Analysis}\label{sec:exp}
\input{sections/experimental_analysis}

\section{Conclusions and Future Work}\label{sec:conclusions}
\input{sections/conclusion}

\section*{
Acknowledgements}
\vspace{-0.2cm}
This work has been partially funded by the European Union’s Horizon 2020 research and innovation programme under the TETRAMAX grant agreement no 761349.

\bibliographystyle{splncs04}
\bibliography{myreferences}

\newpage

\section*{Appendix A}\label{sec:appendix_model}
\input{sections/AppendixModel}

\section*{Appendix B}\label{sec:appendix_experiments}
\input{sections/AppendixExperiments}

\end{document}

%% file: sections/abstract.tex
Artificial Intelligence (AI) and Deep Learning (DL) algorithms are currently applied to a wide range of products and solutions. DL training jobs are highly resource demanding and they experience great benefits when exploiting AI accelerators (e.g., GPUs). However, the effective management of GPU-powered clusters comes with great challenges. Among these, efficient scheduling and resource allocation solutions are crucial to maximize performance and minimize Data Centers operational costs. In this paper we propose ANDREAS, an advanced scheduling solution that tackles these problems jointly, aiming at optimizing DL training runtime workloads and their energy consumption in accelerated clusters. Experiments based on simulation demostrate that we can achieve a cost reduction between 30 and 62\% on average with respect to first-principle methods while the validation on a real cluster shows a worst case deviation below 13\% between actual and predicted costs, proving the effectiveness of ANDREAS solution in practical scenarios.

%% file: sections/intro.tex

Today, artificial intelligence (AI) and deep learning (DL) methods are exploited in a wide range of products and solutions. DL models are usually trained on AI accelerators (e.g., GPU-powered clusters) achieving speed-up factors in the range of 5-40x compared to the CPU-only counterpart~\cite{Intro07}.

The ability to optimize the infrastructure utilization and process the workload with high efficiency under power constraints is critical in this context.  
Indeed, local operators/data centers/Cloud Service Providers (CSPs) are subject to power consumption quotas. Consequently, their financial results are strongly dependent on how efficiently they exploit the infrastructure.  
While there are advanced solutions to manage virtual workloads (viz. virtual machines and containers)~\cite{bernstein2014containers}, the ever-growing DL trainig workload demand on GPUs represents a business opportunity for cloud and data centers operators. However, optimizing revenue requires keeping power consumption under control while maximizing the effectiveness of high-value assets like GPU-powered systems.
The ability to manage the energy footprint in the design of new infrastructure, like distributed computing with GPUs located at the edge in 5G networks, is critical to telecommunication companies owning their infrastructure, to service companies managing infrastructure, and to SMEs developing and providing innovative solutions~\cite{Mishra2018EnergyEfficientDO}. 

Despite the clear advantages of GPU-accelerated clusters in terms of performance, these systems are characterized by a large capital and operating expenditures and a significant energy footprint \cite{DCefficiency,amaral2017topology}.

This paper introduces ANDREAS, an advanced scheduling solution aimed at optimizing  DL training run-time workloads and their energy consumption in accelerated clusters. Our framework, working in an online setting where jobs are expected to be continuously submitted by users in the form of Docker images, implements both a profiling system able to estimate the job processing time and a Random Greedy algorithm able to efficiently and effectively assign at run-time the GPUs of a cluster to the jobs considering energy costs and penalties due to delays in job completions. 
We compare our solution with first-principle approaches commonly used in online scheduling. 
The experiments performed in a simulated environment show that our method guarantees a cost reduction between 30 and 62\% on average with respect to first-principle methods, while the validation on a real cluster highlights a worst case deviation between real and predicted costs below 13\%, proving the effectiveness of ANDREAS solution in practical scenarios.

The remainder of this paper is organized as follows. 
In \autoref{sec:related}, other literature proposals are reviewed. 
Section \ref{sec:architecture} introduces the main elements of the ANDREAS architecture while \autoref{sec:optimizer} formally defines the  problem and describes in detail the proposed algorithm. 
Section  \ref{sec:exp} compares  ANDREAS with first-principle methods. Conclusions
 are finally drawn in \autoref{sec:conclusions}.

%% file: sections/related.tex

The effective management of GPU-accelerated clusters comes with great challenges. Among these,  designing efficient scheduling and resource allocation solutions is crucial to maximize performance and minimize operational costs of data centers. Even if an increasing effort has been put in recent years in tackling both problems, they are often addressed separately, focusing either on jobs scheduling or on resource allocation. In this section, we briefly review previous work in both scenarios.

A resource allocation strategy for DL training jobs is proposed in~\cite{saxena2020effective}. Authors determine the optimal jobs batch size  according to their scaling efficiency and combine a FIFO scheduling with a dynamic programming algorithm to determine an effective solution.  Similarly, \textit{Optimus}~\cite{peng2018} is a Kubernetes scheduler designed to manage DL jobs on a shared distributed containerized environment, which minimizes jobs training time by relying on online resource-performance models. $DL^2$~\cite{Peng2019DL2AD}, also based on Kubernetes, dynamically allocates resources by exploiting Deep Reinforcement Learning instead of relying on analytical performance models and job profiling.

On the scheduling side, \textit{Gandiva}~\cite{xiao2018gandiva} focuses on tasks recurrent behaviours during the sequence of mini-batches iterations and proposes a preemptive, time-sharing scheduler for DL jobs.

GPU cluster managers closer to ANDREAS are \textit{Tiresias}~\cite{tiresias}, \textit{Philly}~\cite{philly}, and \textit{FfDL}~\cite{FfDL}, that implement integrated frameworks supporting the full training process and dealing with jobs submissions and execution. However, they focus only on the scheduling of DL training jobs,  while the resources to be assigned to all applications are specified by the users.

%% file: sections/architecture.tex

\begin{figure*}[h!t]
\centering
\includegraphics[width=0.8\textwidth]{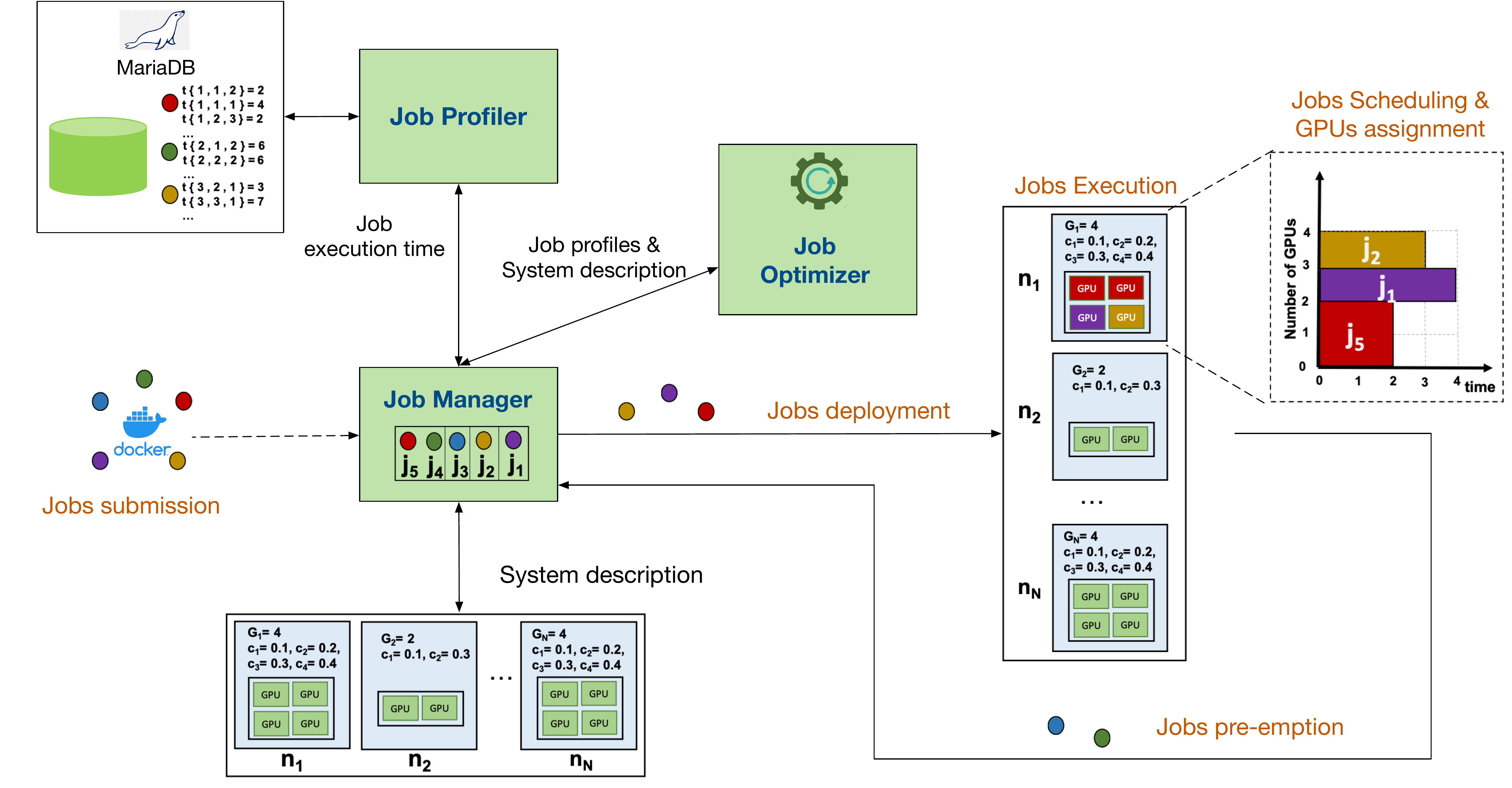}
\caption{ANDREAS framework. \footnotesize{In the example, we consider five submitted jobs $j_1, j_2, j_3, j_4, j_5$. 
Nodes $n_1$ to $n_N$ are equipped with a variable number of GPUs, of possibly different type, and have different energy costs. Jobs $j_1$, $j_2$ and $j_5$ are deployed on $n_1$, running on 1 GPU and 2 GPUs, respectively. The other jobs are sent back to the queue and no other nodes are selected.}}
\label{fig:framework}
\end{figure*}

The high-level architecture of the ANDREAS framework is outlined in \autoref{fig:framework}. ANDREAS includes three main components: the \textit{Job Manager}, the \textit{Job Profiler} and the \textit{Job Optimizer}. 
Training applications are submitted by users in form of Docker images.
The Job Manager receives the users' requests and orchestrates the execution of the submitted jobs as follows. First of all, the Job Profiler is invoked to collect information about the expected per-epoch execution time of the submitted jobs on the different resources available in the system. One node (characterized by a given type and number of GPUs) is dedicated to collect these data, which are then stored in  MariaDB. 
If the cluster is heterogeneous (i.e., it includes nodes with different CPU or GPU types), multiple nodes can be engaged in profiling activities; the profiling configuration can even be changed dynamically while production applications run.

The collected profiling data, together with a description of the system, with all available nodes, are provided to the Job Optimizer, which is in charge of selecting the optimal deployment for the submitted jobs. In particular, the Job Optimizer exploits the algorithm described in the following section to determine which resource configuration should be used to deploy and run the different jobs. If the resources available in the system are insufficient to run all jobs concurrently, some of them are preempted, and their execution is postponed. 

The Job Optimizer is invoked by the Job Manager periodically, or in reaction to re-scheduling events (viz., a new job submission or completion). The Job Manager deploys the jobs according to the optimization results, migrating the ones in progress among different GPUs if necessary. 

%% file: sections/model.tex
This section  presents the main ANDREAS assumptions    and describes the problem; then, it focuses on the Randomized Greedy algorithm we propose to solve it.

\subsection{Assumptions and Problem Description}

This section addresses the capacity allocation problem solved by the Job Optimizer at every rescheduling point along with the main underpinning assumptions. 
The viewpoint of a CSP running a data center is considered.
The data center is assumed to be composed by a set of heterogeneous nodes  denoted by $\mathcal{N} = \{n_1, n_2, ..., n_N\}$, with $N = |\mathcal{N}|$.
Each node is characterized by a specific type and number of available GPUs. The number of GPUs available on node $n$ is identified by a set $\mathcal{G}_n = \{1, 2, ..., G_n\}$. The node is characterized also by a cost per unit time $c_{ng}$, when $g$ GPUs are in use. GPU homogeneity, however, is not preserved across different nodes.

Jobs are submitted continuously over time by several users, are profiled and executed by  ANDREAS  on a suitable node.  
Each node $n \in \mathcal{N}$ can host multiple virtualized workloads, assigning in a dedicated fashion the available GPUs to jobs; a job, in turn, runs on a single node. In addition, since the number of available resources is limited, jobs that cannot be accommodated in the system are entered into a waiting set until the next rescheduling point.  Let $\mathcal{J}$ represent the set of submitted jobs. Each job $j\in \mathcal{J}$ is characterized by a due date $d_j$, a tardiness penalty $\omega_j$, and a penalty to pay for postponing a job $\rho$.  
Furthermore, each job is associated with an execution time $t_{jng}$ that depends on the job, the number of GPUs $g$ and their type, which in turn depends on the considered node  $n$.
The parameters and variables of the problem that are considered are reported in \autoref{tab:MS_PBdescr}.

\input{tables_and_algorithms/MStable}

For each rescheduling point, the problem to be solved is a capacity allocation problem where for each job $j \in \mathcal{J}$ it must be decided whether to execute it right away or to postpone it. In the former case, the system also has to determine in which node to host the job and the resources (GPUs) to assign to its execution. If jobs are preempted, they will restart from the last snapshot (under the assumption that a snapshot of the DL model weight is taken every few epochs).

The primary objective of ANDREAS is to minimize costs, 
including tardiness-caused penalties and energy related execution costs (including, e.g., air conditioning, UPS overhead, etc.). Nonetheless, the Job Optimizer at each rescheduling point considers a problem with a much narrower scope making it challenging to estimate the impact of decisions on the overall cost. 
For this reason, the following proxy function, denoted by $f_{\text{OBJ}}$, is considered:

\begin{footnotesize}
	\begin{equation}
	\min \left[\sum_{j \in \mathcal{J}}\left(\omega_j\tau_j + \rho\omega_j\hat{\tau}_j\right) + 
	\sum_{j \in \mathcal{J}, n \in \mathcal{N}}\alpha_{jn}\pi_{jn}\right]
	\label{eq:MS_ObjFunc}
	\end{equation}
\end{footnotesize}

In particular, the proxy is expressed by two terms: the first 
one represents the job tardiness and the worst-case tardiness penalty costs (where $\hat{\tau}_j$ is the tardiness a job may incur in if it is postponed to the next period). 
The second term, in turn, takes into account operation costs of the fastest jobs on all nodes; $\alpha_{jn}$ is 1 if job $j$ is the first-ending job on node $n$ while $\pi_{jn}$ represents its execution cost.
This function has been designed with the precise aim of penalizing configurations with delayed jobs and, at the same time, incentivizing solutions with short-duration jobs. In this way, ANDREAS can provide a lower average job latency. 
In addition, since the Job Optimizer is invoked every time the fastest job finishes its execution, such a proxy function tends to promote higher system responsiveness and better control over the quality of the overall schedule.  

For a matter of space, the complete Mixed Integer NonLinear Programming (MINLP) formulation of the problem is reported in  Appendix A.

%% file: tables_and_algorithms/MStable.tex

\begin{table}[t]
	\caption{Problem parameters and variables}
	\footnotesize
	\label{tab:MS_PBdescr}
	\begin{tabular*}{\columnwidth}{| l | l |} 
		\hline
		\multicolumn{2}{| l |}{\textbf{Parameters}}\\   
		\hline
		$\mathcal{J}$          & set of submitted jobs \\
		$\mathcal{N}$          & set of available nodes \\
		$\mathcal{G}_n$           & set of available GPUs in each node $n \in \mathcal{N}$\\
		$G_n$           			& number of available GPUs in each node $n \in \mathcal{N}$\\
		$c_{ng}$       & energy cost of each node $n \in \mathcal{N}$ when $g$ GPUs are used \\
		$d_j$       & due date of job $j \in \mathcal{J}$ \\
		$\omega_j$  & tardiness weight of job $j \in \mathcal{J}$ \\
		 $t_{jng}$    & execution time of job $j \in \mathcal{J}$ when running on $g$ GPUs on \\
		 & node $n \in \mathcal{N}$\\
		$\rho$             & penalty coefficient for postponed jobs\\
		
		\hline
		\multicolumn{2}{| l |}{\textbf{Variables}}\\    
		\hline
		$w_n$       & 1 if node $n \in \mathcal{N}$ is chosen \\
		$\tau_j$    & tardiness of job $j \in \mathcal{J}$ \\
		$\hat{\tau}_j$    & worst-case tardiness of job $j \in \mathcal{J}$ \\
		$\pi_{jn}$  & operation cost of job $j \in \mathcal{J}$ on node $n \in \mathcal{N}$ \\
		$\alpha_{jn}$  & 1 if job $j \in \mathcal{J}$ is the first ending job on node 
		$n \in \mathcal{N}$ \\
		\hline
	\end{tabular*}
	\vspace{-0.2cm}
\end{table}

%% file: sections/algorithm.tex

\subsection{ANDREAS Randomized Greedy algorithm}

The randomized greedy heuristic that we have developed for the capacity allocation problem is based on the following criteria and assumptions.
\begin{itemize}
   \item Jobs are selected according to their pressure, which is an index that measures how close they are to the due date when executed with the fastest configuration. For each job $j \in \mathcal{J}$, the pressure $\Delta_j$ is:
\begin{footnotesize}
    \begin{equation}\label{eq:pressure}
        \Delta_j = T_c + \min_{\substack{n \in \mathcal{N}\\ g \in \mathcal{G}_n}}{\left\{t_{jng}\right\}} - d_j,
    \end{equation}
    \end{footnotesize}
    where $T_c$ denotes the current scheduling time. 
    
    \item The best configuration $\left(n,g\right) \in \mathcal{N} \times \mathcal{G}_n$ for each selected job is either (i) the cheapest configuration such that the job is executed before its due date, if such a configuration exists, or (ii) the fastest available configuration if, independently from the selected setup, it is not possible to execute the job before its due date. Since in our framework the time unit execution cost of a job on any available configuration is always lower than the penalty incurred if the job due date is violated, it is reasonable to choose the cheapest configuration as long as the due date can be met, while, if it is violated, it is more convenient to complete the job execution as fast as possible to reduce the corresponding penalty.
    
    \item Execution costs increase linearly with the number of GPUs, as demonstrated by GPU-based servers energy profiles \cite{FfDL}, while the speedup of the processing time is sublinear in the number of GPUs, as observed in GPU-based application benchmarks~\cite{closer19}.
\end{itemize}

The randomized greedy algorithm, which is summarized in \autoref{lst:RG}, receives as input the set of jobs $\mathcal{J}$ in the queue, their pressures and a maximum number of iterations, and returns a solution specifying for every job 
whether to execute it and the corresponding configuration.
At each iteration, a candidate solution $S$ is constructed via randomized greedy choices and is stored in $S_{best}$ if $f_{\text{OBJ}}(S) < f_{\text{OBJ}}(S_{best})$.

\input{tables_and_algorithms/RG}

At line 6, the set $\mathcal{J}$ of submitted jobs is sorted, producing as output a new set denoted by $\mathcal{J}_s$. Note that job $j$ precedes job $k$ in $\mathcal{J}_s$ if and only if $\Delta_j > \Delta_k$, i.e., job $j$ is more likely to violate its due date. As a first randomization step, some jobs in $\mathcal{J}_s$ can be swapped, with probability inversely proportional to their tardiness weight.

For all jobs $j$ in the sorted list $\mathcal{J}_s$, the set of configurations (i.e., pairs made by a node and a certain number of GPUs) such that the job can be executed within its due date (determined by the condition $t_{jng} + T_c < d_j$) is defined, in line 8, as:

\begin{footnotesize}
\[
	D^*_j = \left\{(n,g) \in \mathcal{N} \times \mathcal{G}_n 
						:~ T_c + t_{jng} < d_j\right\}.
\]
\end{footnotesize}
The set $D^*_j$ is used to select the best configuration for job $j$, according to the following rules. Since one of the main contributions to the total cost of the schedule is the penalty for due date violations, we always try to select configurations that guarantee to complete the jobs without tardiness (or with the least possible delay, if there is no possibility to match the required due date). Therefore, we define the optimal configuration as:

\begin{footnotesize}
\begin{equation*}
    \left(n^*, g^*\right) = 
    \begin{cases}
        \argmin_{D^*_j}\left\{t_{jng}c_{ng}\right\}  &\text{ if } D^*_j \neq \emptyset \\
        \argmin_{n \in \mathcal{N}, g \in \mathcal{G}_n}\left\{t_{jng}\right\} &\text{ otherwise.}
    \end{cases}
\end{equation*}
\end{footnotesize}

In the first case, the set $D^*_j$ is not empty, which entails that there exist configurations $(n,g)$ such that the job can be executed before its due date. In this case, $(n^*,g^*)$ is identified by the cheapest configuration, i.e., by the element in $D^*_j$ with minimum cost (identified by the product between the time-unit cost $c_{ng}$ and the expected execution time $t_{jng}$). In the second case, the job due date is violated with any available configuration, therefore we select the one that guarantees the minimum expected execution time, in order to reduce the tardiness as much as possible.
The algorithm explores the space of possible configuration by introducing some randomness in the the configuration selection process, where, instead of choosing the configuration according to the rule defined above, it selects as candidate configuration for job $j$ one of the $(n, g)$ with lower cost, with probability inversely proportional to the cost itself. Having denoted by $(n^*,g^*)$ the selected configuration for job $j$, the assignment proceeds as follows:

\begin{enumerate}
    \item The algorithm tries to assign job $j$ to the required node $n^*$, provided that the number of available GPUs in $n^*$ is enough to assign $g^*$ of them to job $j$.
    
    \item If this assignment is not feasible, it tries to assign job $j$ to a suboptimal configuration available in an open node. Iteratively, it loops over the elements in $D^*_j$ until it finds a configuration that fits in any node $n' \in \mathcal{N}_O$.
\end{enumerate}

%% file: tables_and_algorithms/RG.tex

\begin{algorithm}[t]
\begin{footnotesize}
	\caption{Randomized construction procedure}
	\label{lst:RG}
	\begin{algorithmic}[1]
		\Function{\Call{randomized\_construction}{$\mathcal{J}$, $\Delta$, $\text{MaxIt}_{\text{RG}}$}}{}
		    \State iter = 0
    		\State $S_{\text{best}} \gets $ empty schedule
    		\While {iter $<$ $\text{MaxIt}_{\text{RG}}$} 
    		    \State $S \gets$ empty schedule   \Comment{$S$: current schedule}
    		    \State $\mathcal{J}_s \gets$ \Call{sort\_jobs\_list}{$\mathcal{J}$, $\Delta$}\label{sort_queue}		\Comment{$\Delta: $ pressures of all jobs}
        		\ForAll {$j \in \mathcal{J}_s$}	\Comment{$\mathcal{J}_s: $ sorted queue}
            		\State $D^*_j = \left\{(n,g) \text{ s.t. } t_{jng} + T_c < d_j\right\}$\label{dstar}
            		\State $(n^*, g^*) \gets$ \Call{select\_best\_configuration}{$j, D^*_j$}
            		\State assigned $\gets$ \Call{assign}{$j, (n^*, g^*), \mathcal{N}_O$}
            		\If {not assigned}
                		\State \Call{assign\_to\_suboptimal}{$j, S, \mathcal{N}_O$}
                	\Else
                	    \State $S \gets S \cup \left(j, n^*, g^*\right)$
            		\EndIf
            		\State $\mathcal{J}_s \gets \mathcal{J}_s \setminus \{j\}$
        		\EndFor
        		\If{$f_{\text{OBJ}}(S) < f_{\text{OBJ}}(S_{\text{best}})$}
        		    \State $S_{\text{best}} \gets S$
        		\EndIf
        		\State iter $\gets$ iter + 1
    		\EndWhile
    		\State \textbf{return} $S_{\text{best}}$
		\EndFunction
	\end{algorithmic}
	\end{footnotesize}
\end{algorithm}

%% file: sections/experimental_analysis.tex

This section describes the results obtained through an experimental campaign, aiming at assessing the effectiveness of ANDREAS framework in real-case scenarios.

\input{sections/exp_setup}

\subsection{Profiling Analysis}\label{sec:exp_profiling}
\input{sections/exp_profiling}

\subsection{Simulations}\label{sec:exp_simulations}
\input{sections/exp_simulations}

\subsection{Validation in the ARMIDA setup}\label{sec:exp_real}
\input{sections/exp_real}

%% file: sections/exp_setup.tex

The experimental campaign involved heterogeneous Deep Learning jobs based on different network types, namely multi-layer LSTM~\cite{lstm}, EfficientNet~\cite{efficientNet}, and simple Convolutional Neural Networks~\cite{convNet}, referred to in the following as ConvolutionNet. All the jobs were implemented relying on TensorFlow 2 and during the training model weights snapshots are taken every epoch. 
The Job Optimizer has been configured with $\text{MaxIt}_{\text{RG}}=1,000$
All experiments described in the following sections have been performed relying on ARMIDA (ARM Infrastructure for the Development of Applications) infrastructure, a test \& development 8-node cluster located at E4’s premises. ARMIDA configuration is detailed in Appendix B. The types and numbers of GPUs available are summarized in \autoref{tab:catalog}.

\input{tables_and_algorithms/armida_nodes}

%% file: tables_and_algorithms/armida_nodes.tex

\begin{table}[ht]
    \centering
    \begin{scriptsize}
    \caption{Configurations available in ARMIDA}
    \label{tab:catalog}
    \centering
    \begin{tabular}{|l|l|r|}
        \hline
        \textbf{Node ID}    & \textbf{GPU type}   &  \textbf{\# GPU}\\ \hline
        armida-04   & Nvidia Tesla V100-PCIE 32GB   & 1\\
        armida-05   & Nvidia Tesla V100-PCIE 32GB   & 1\\
        armida-06   & Nvidia Tesla V100-PCIE 32GB   & 2\\
        armida-07   & Nvidia Tesla T4 (TURING)  & 1\\
        \hline
    \end{tabular}
    \end{scriptsize}
    \vspace{-0.2cm}
\end{table}

%% file: sections/exp_profiling.tex

The profiling analysis described in this section has been performed to measure the power consumption of ARMIDA nodes and GPUs, as well as the expected execution times of the considered applications on all the configurations available in the reference hardware system. The profiling has been performed through  ExaMon (Exascale Monitoring) a highly scalable framework for the performance and energy monitoring of HPC servers~\cite{examon2}.

The power consumption of ARMIDA nodes allowed to compute the hourly energy costs (starting from the base cost of 0.172 EUR/KWh) used by the optimizer to determine the best assignment for the different jobs, according to the different configurations. To evaluate also the energy consumption due to the air conditioning system, the UPS and other overheads of a data center environment, the energy costs has been inflated by considering a PUE (Power Usage Effectiveness)~\cite{DCefficiency} equal to 1.33, which is the value measured in the ARMIDA setup.

%% file: sections/exp_simulations.tex

The results that can be achieved by exploiting the ANDREAS optimizer at a larger scale have been evaluated through a set of simulations, comparing the outcome of the Randomized Greedy (RG) algorithm against first-principle methods as First In First Out (FIFO), Earliest Deadline First (EDF) and Priority Scheduling (PS), as it is done in other literature proposals~\cite{amaral2017topology}.

Different instances of the experiments have been generated by randomly selecting a set of DL jobs with different characteristics in terms of batch sizes and number of required training epochs. We have considered multiple scenarios, involving a number $N$ of nodes varying from 10 to 100, and a number of jobs $J = 10N$. Jobs inter-arrivals  are characterized by a \textit{mixed} rate, generated as described in~\cite{saxena2020effective}. The characteristics of nodes are similar to those reported in \autoref{tab:catalog}. Specifically, half of them are equipped with GPUs of type TeslaV100, while the others with GPUs of type TURING. We have considered two different cases: in the first one, nodes are characterized by two TeslaV100 GPUs  or by one TURING; in the second case nodes  with four TeslaV100 or two TURING GPUs are considered. The parameter $\rho$ for penalizing jobs postponements has been set to 100.

The results of the RG algorithm and of FIFO, EDF and PS methods in the first scenario are reported in \autoref{fig:costs_and_times_2-1}. 
The first plot reports the expected energy costs of jobs execution, while the second shows the total cost, which is the sum of the energy cost plus the penalties due to jobs’ due dates violations. A significant reduction, of around 62\% on average, is obtained considering the total costs. Indeed, first-principle methods never change the configuration assigned to a job once it has been started, which entails that, if jobs with higher priority are submitted after long-running jobs with lower priority, they have to wait for them to be completed before receiving resources. 
The third plot shows the expected time required to completely execute the training of all the $J$ jobs submitted to the system. The reduction guaranteed by the RG algorithm helps in minimizing the probability that jobs due dates are violated, consequently reducing the corresponding penalty costs.
Finally, the last plot reports the time required to the optimizer to solve one instance of the problem, answering the call of the Job Manager providing the schedule of all jobs. Even if our RG algorithm requires a longer time than first-principle methods to provide a solution, this time remains always below 0.1 seconds, which is significantly lower than the time required to reconfigure the system.

The results of the RG algorithm and of FIFO, EDF and PS methods in the second scenario are reported in \autoref{fig:costs_and_times_4-2}. Again, it is possible to notice that the average cost reduction guaranteed by the RG algorithm is of almost 30\%, while the time required to determine a solution is lower than 0.01 seconds on average.

\input{figures/fig_costs}

%% file: figures/fig_costs.tex

\begin{figure*}[ht]
	\centering
	\begin{minipage}[l]{1.0\columnwidth}
		\includegraphics[width=.48\linewidth]{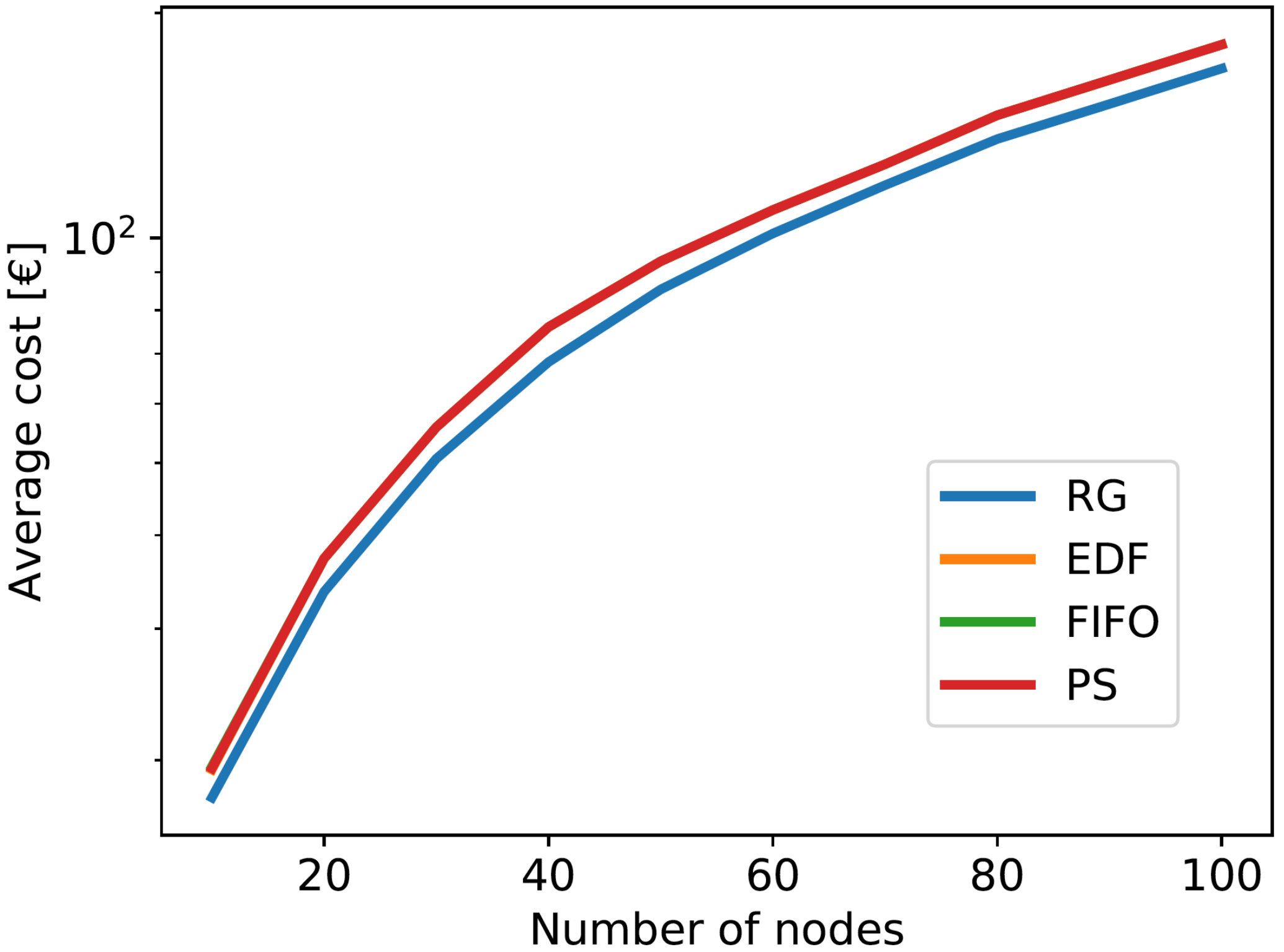}
		\hfill
		\includegraphics[width=.47\linewidth]{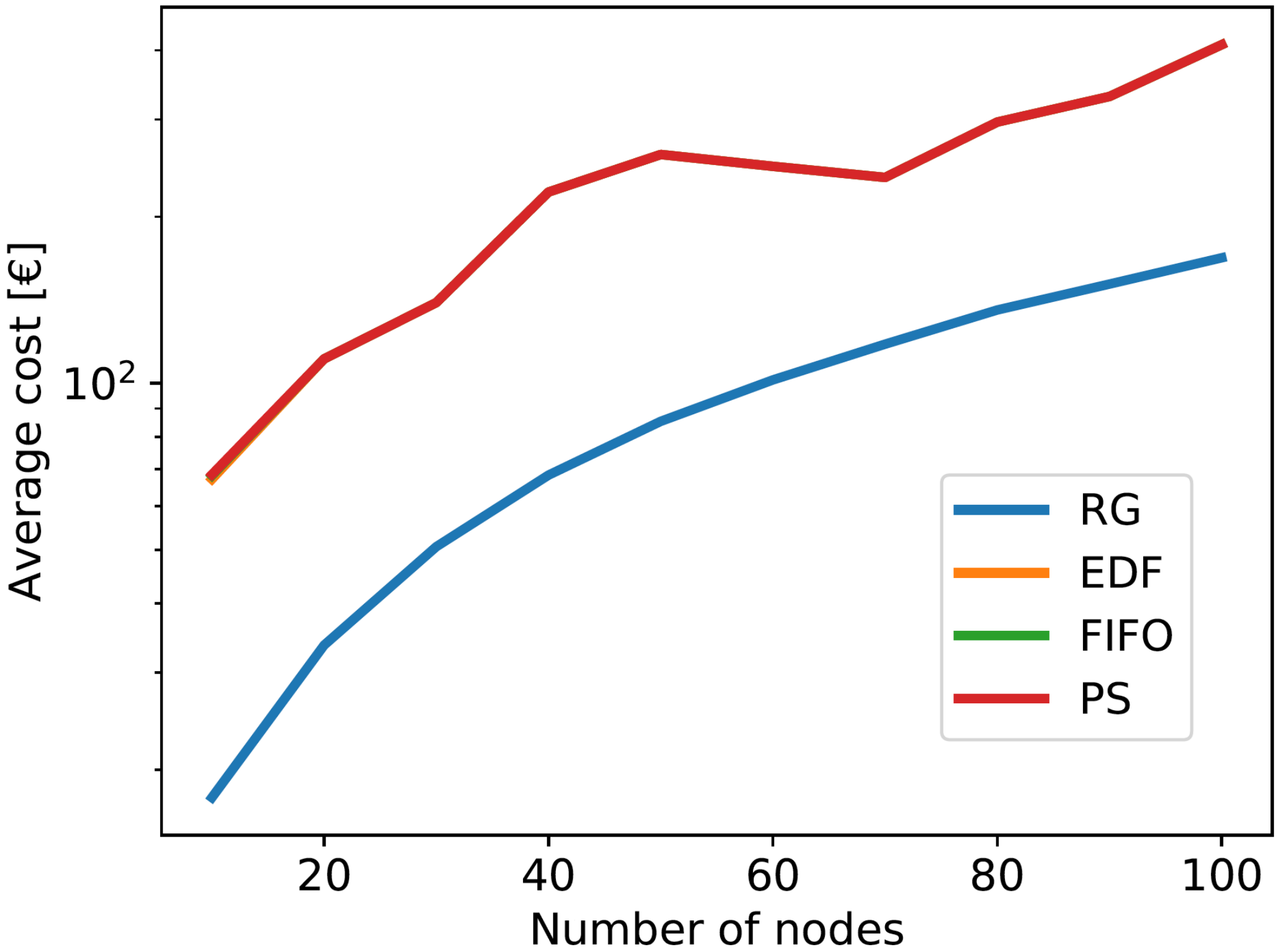}
		\subcaption{Energy costs and total costs}
		\label{fig:costs42}
	\end{minipage}
	\hfill
	\begin{minipage}[r]{1.0\columnwidth}
		\includegraphics[width=.48\linewidth]{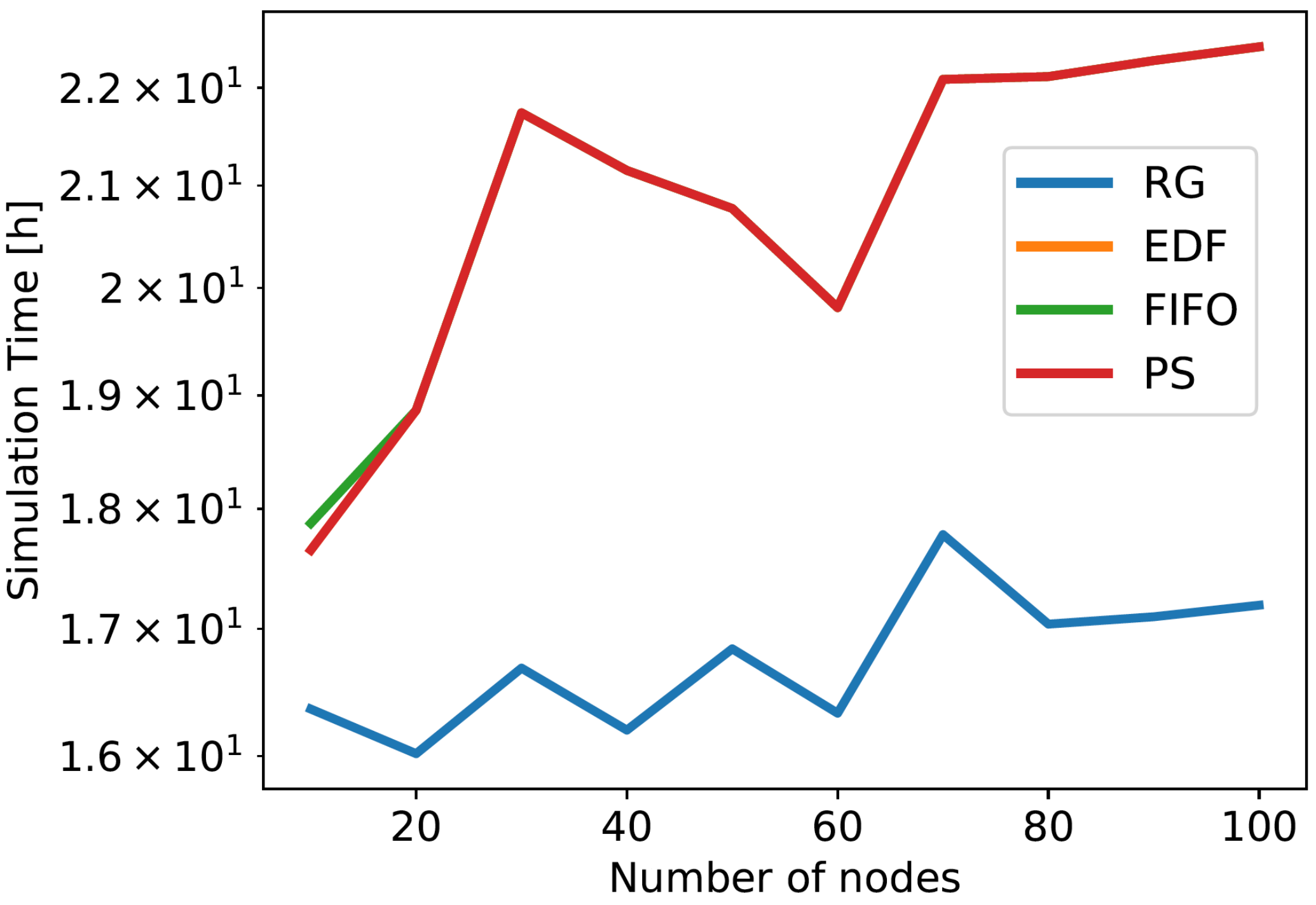}
		\hfill
		\includegraphics[width=.47\linewidth]{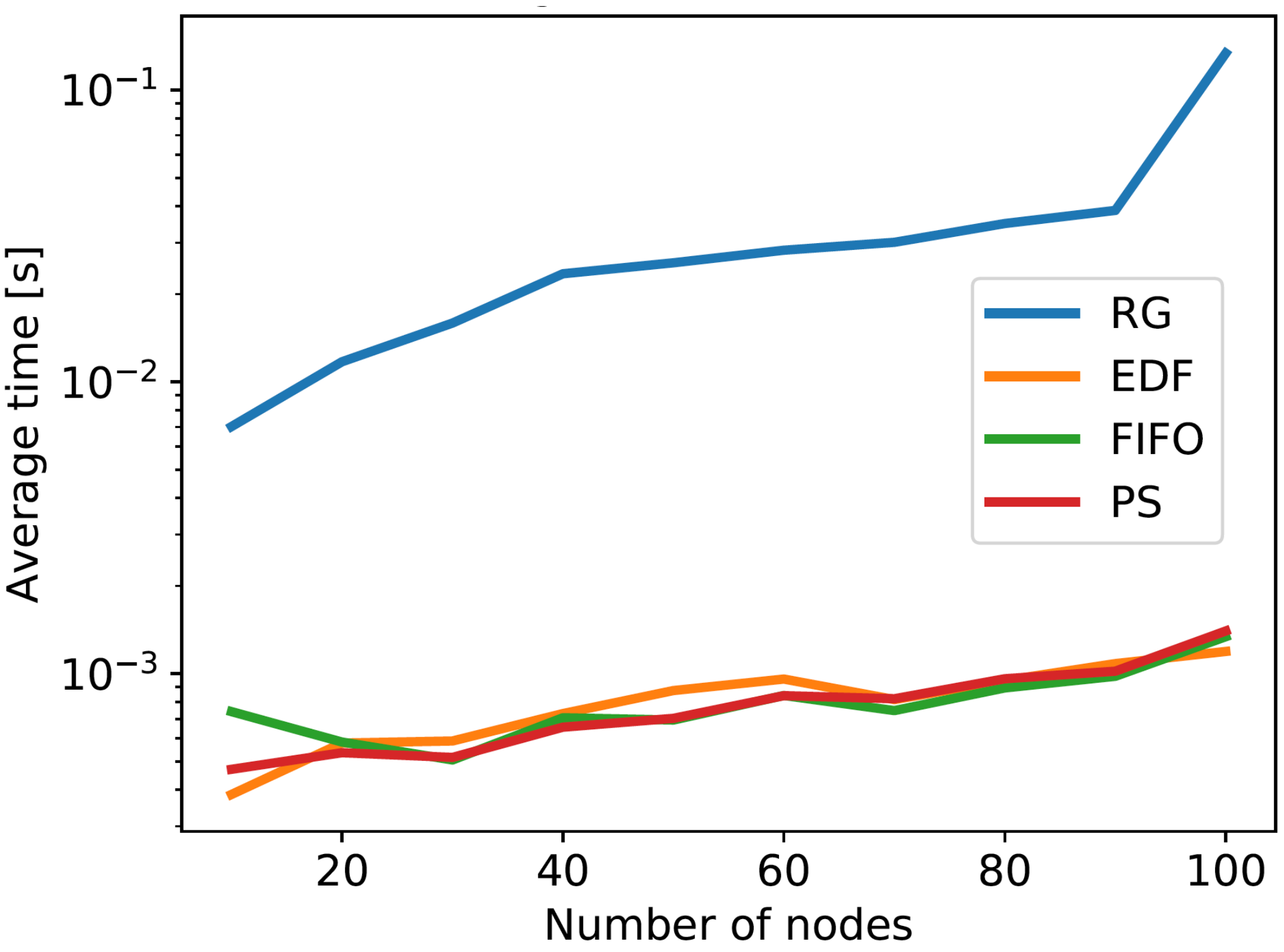}
		\subcaption{Simulation time and avg. execution time per optimizer call}
		\label{fig:times42}
	\end{minipage}
	\caption{Average costs and execution times with the different methods - first scenario}
	\label{fig:costs_and_times_4-2}
\end{figure*}

\begin{figure*}[ht]
	\centering
	\begin{minipage}[l]{1.0\columnwidth}
		\includegraphics[width=.48\linewidth]{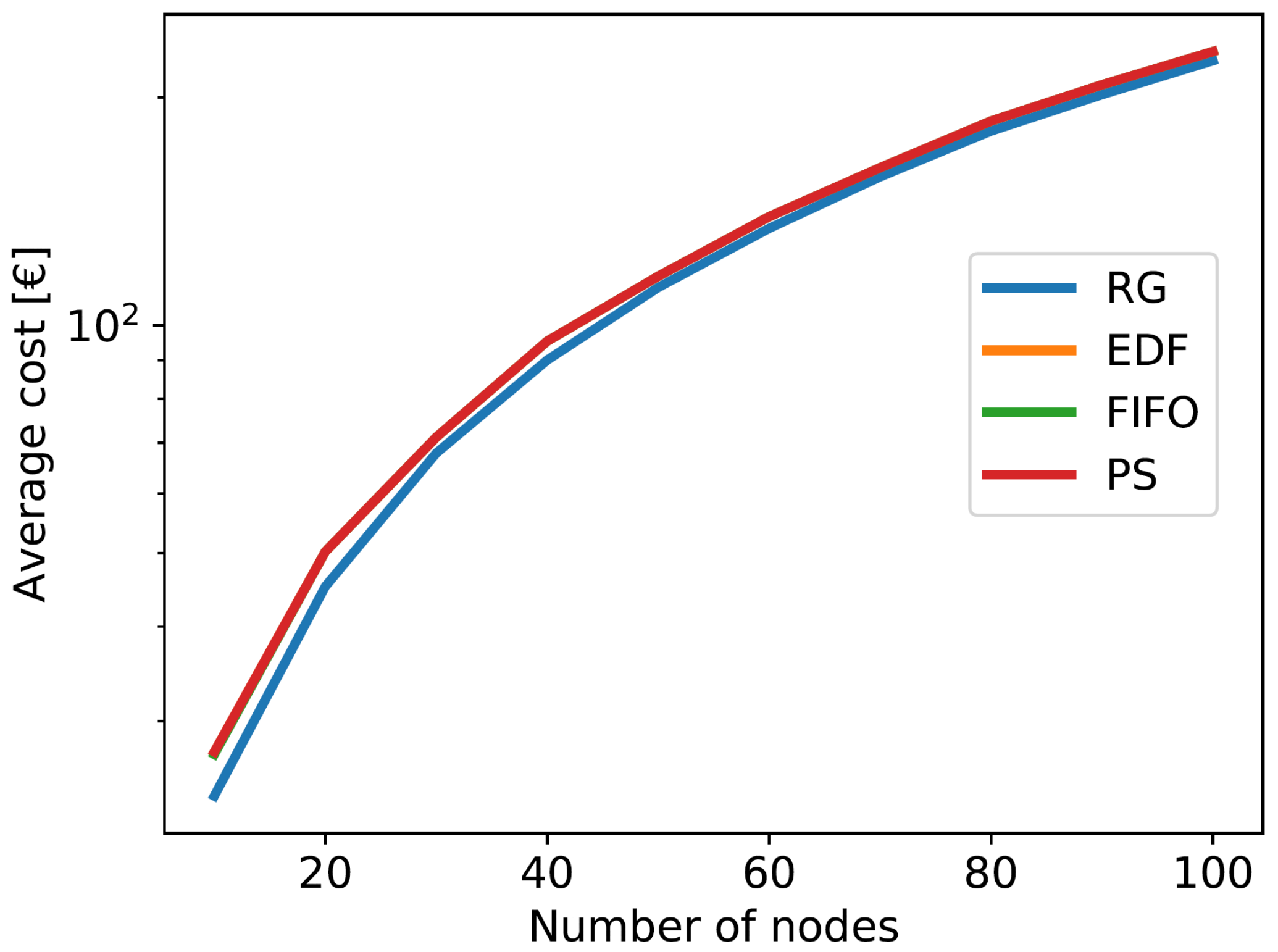}
		\hfill
		\includegraphics[width=.47\linewidth]{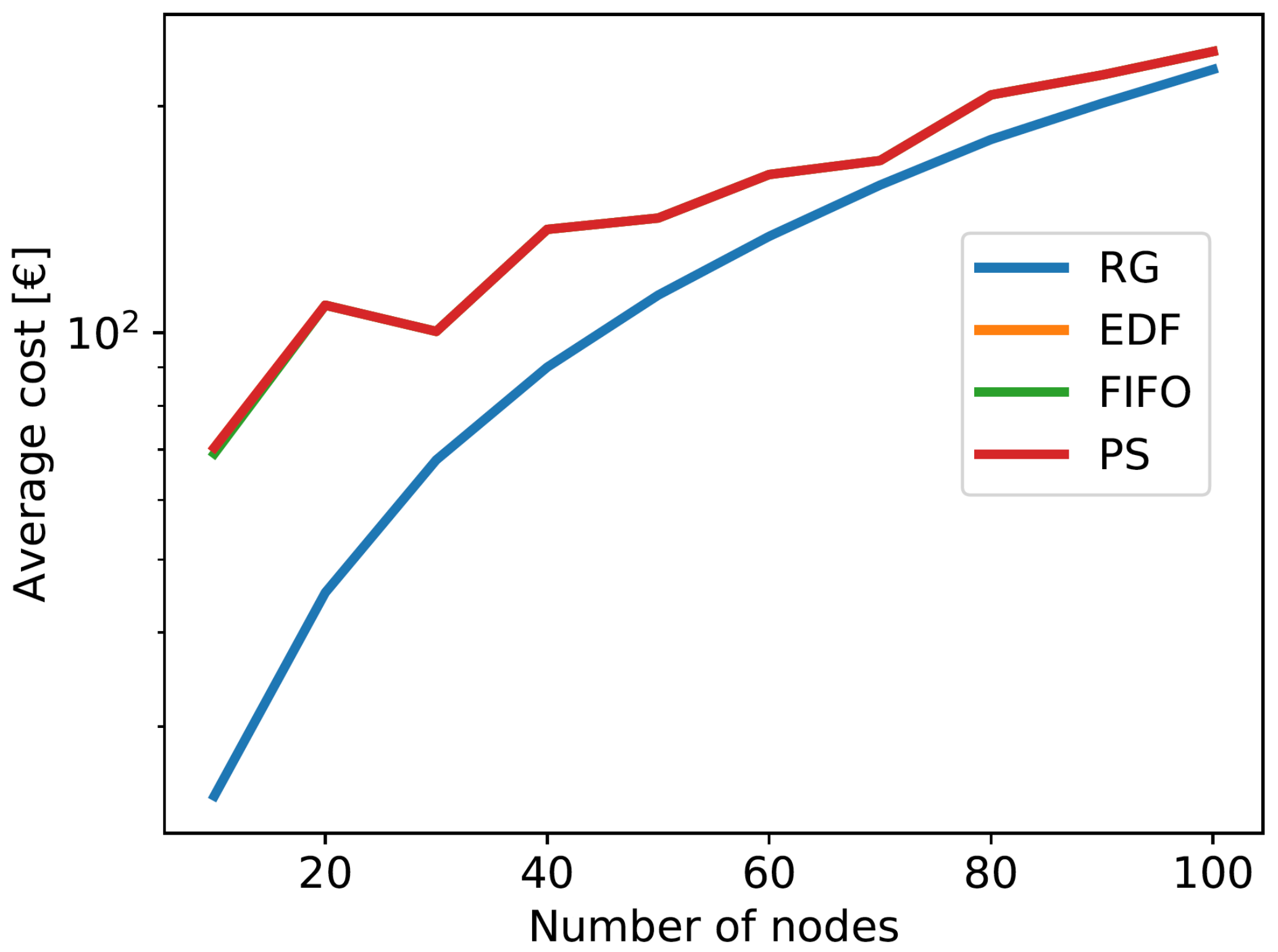}
		\subcaption{Energy costs and total costs}
		\label{fig:costs21}
	\end{minipage}
	\hfill
	\begin{minipage}[r]{1.0\columnwidth}
		\includegraphics[width=.48\linewidth]{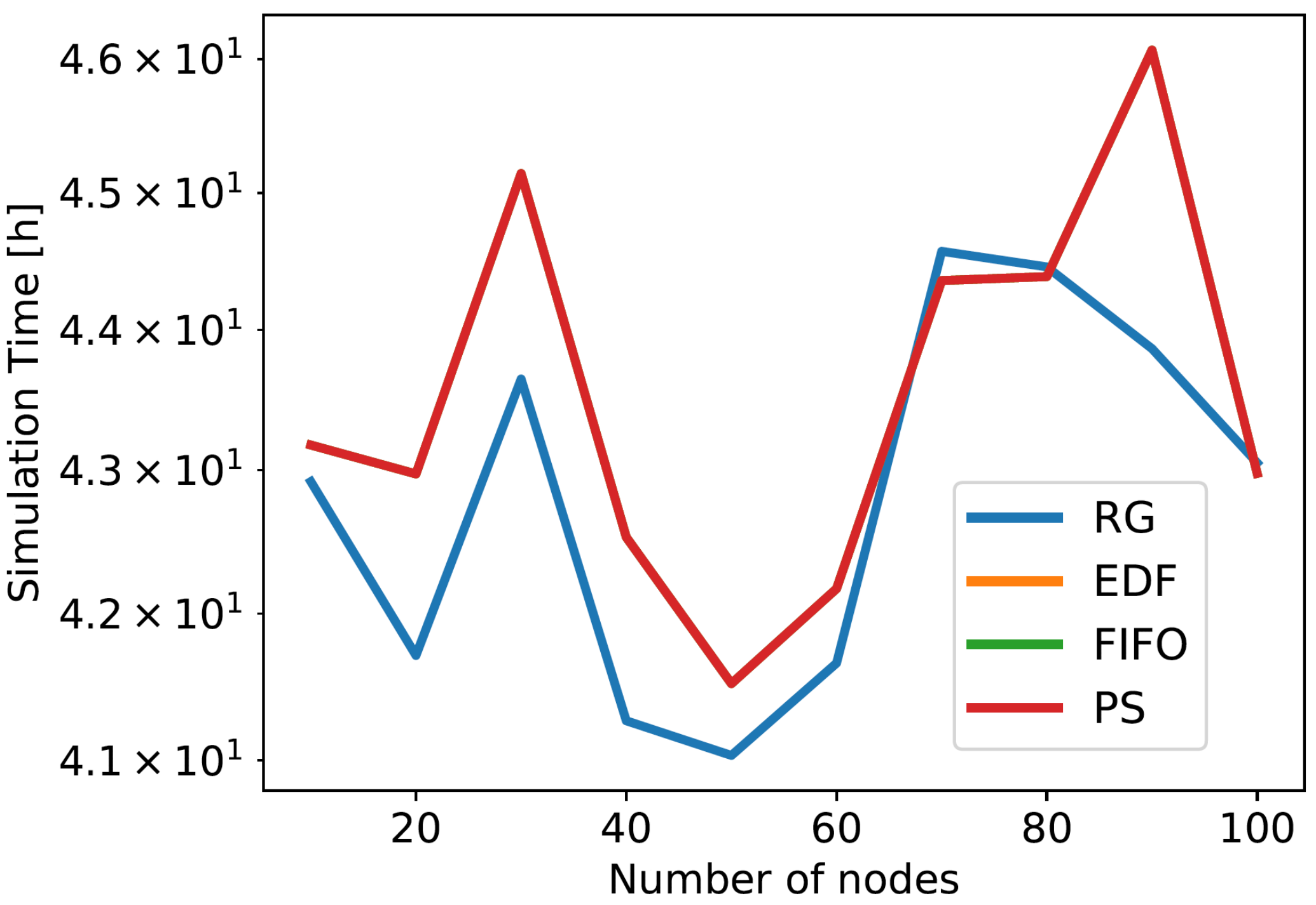}
		\hfill
		\includegraphics[width=.47\linewidth]{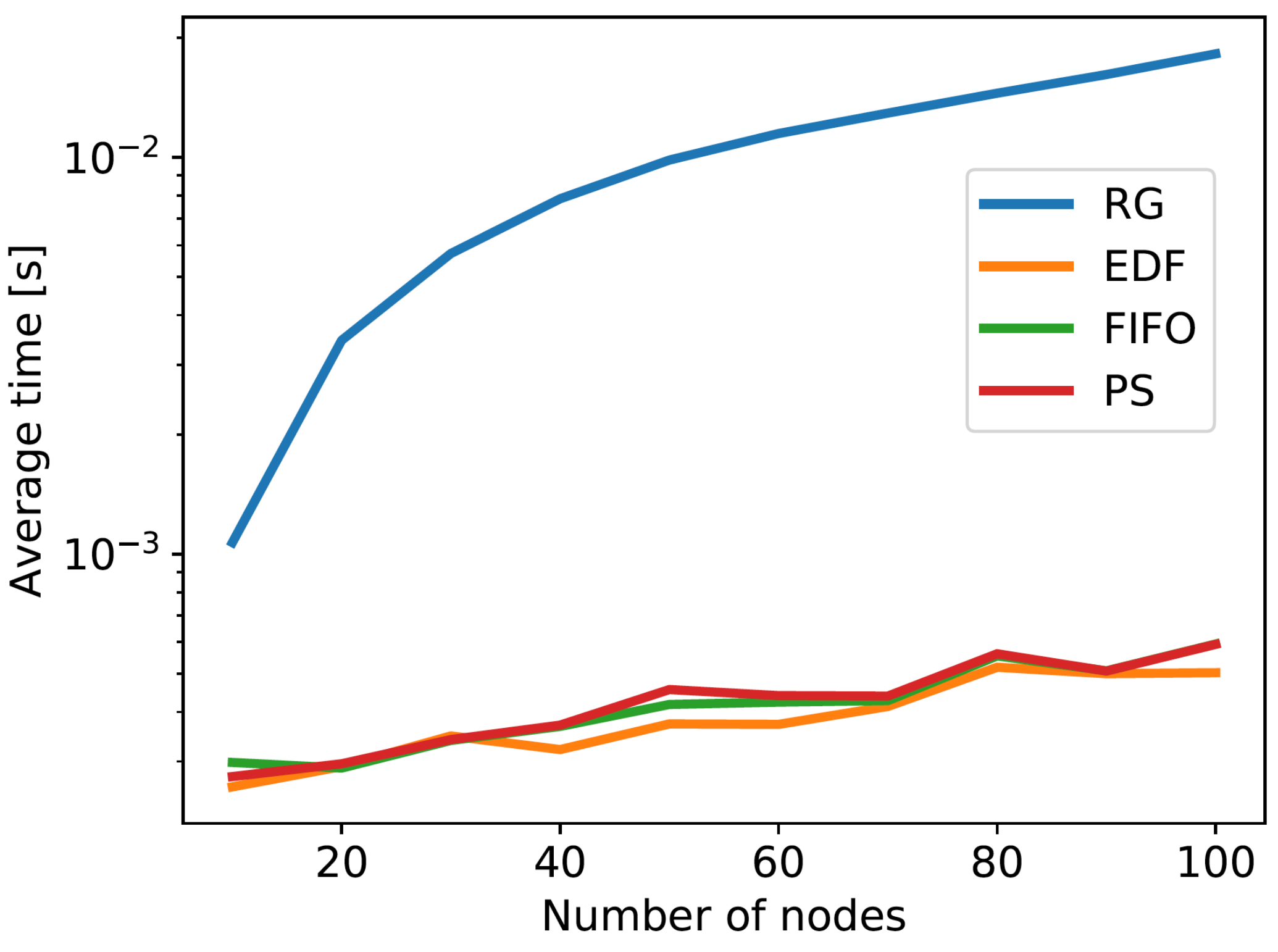}
		\subcaption{Simulation time and avg. execution time per optimizer call}
		\label{fig:times21}
	\end{minipage}
	\caption{Average costs and execution times with the different methods - second scenario}
	\label{fig:costs_and_times_2-1}
\end{figure*}

%% file: sections/exp_real.tex

The results provided by the ANDREAS framework have been validated in the ARMIDA setup by considering a prototype system including 8 different jobs. Their characteristics are summarized in \autoref{tab:jobs} of Appendix B. 
The inter-arrival time has been set to 1200s. This is an accelerated scenario, in terms of submission frequency and average execution times of applications. Nonetheless, this does not affect the results effectiveness since the workload assigned to the available nodes is comparable to practical situations~\cite{saxena2020effective}.

The system is composed by three nodes, which correspond to armida-05, armida-06 and armida-07 as described in \autoref{tab:catalog}; the node armida-04 is not considered, since it is dedicated to profiling. The aim of this experiment was to assess the accuracy of our framework, by measuring the gap between the predicted energy costs and the actual costs measured in the ARMIDA nodes.

The sequence of jobs executions, with the relative configurations, is reported in \autoref{fig:RealTrace} of Appendix B. It is possible to observe both processes that characterize the decisions of the RG method: first of all, GPU sharing, that can be observed for instance when J10 is submitted and it is assigned to armida-06 together with J6. Then, preemption, that comes into place, for instance, when J2 is submitted: since it has a stricter due date than J7, it is immediately deployed on armida-07, while the execution of the latter is postponed to a future scheduling point.

\autoref{fig:RealCosts} in Appendix B reports the power consumption of ARMIDA nodes during the full experiment session. This has been used to compute the actual energy costs incurred by the system during the whole experiment, including the reconfiguration costs that are not considered by our simulator. We have compared this result with our prediction, noticing that, in the accelerated framework we have considered, all jobs are completed within the due dates. 
\autoref{tab:deviations} reports the deviations of the predicted cost with respect to the real cost (the energy cost is equal to the total cost, since no due date violations are observed). It is relevant to notice that the predicted cost is higher than the real cost, which makes our framework more reliable and validates its adoption in an industry setting.

\input{tables_and_algorithms/deviations}

%% file: tables_and_algorithms/deviations.tex

\begin{table}[ht]
    \centering
    \begin{scriptsize}
    \caption{Results of the validation on ARMIDA}
    \label{tab:deviations}
    \centering
    \begin{tabular}{|l|r|}
        \hline
        \textbf{}    & \textbf{Energy cost}\\ \hline
        \textbf{Real cost [€]}   & 0.95111\\
        \textbf{Predicted cost [€]}   & 1.08438\\
        \textbf{Deviation}   & 12.29\%\\
        \textbf{Average deviation for each optimizer call}   & 10.81\%\\
        \hline
    \end{tabular}
    \end{scriptsize}
    \vspace{-0.2cm}
\end{table}

%% file: sections/conclusion.tex

The ANDREAS scheduling solution proposed in this work effectively tackles the scheduling and resource allocation problems in GPU-powered clusters, achieving a cost reduction between 30 and 62\% on average with respect to first-principle methods. The deviation between the predicted and the real costs in a validation cluster is below 13\%, supporting its adoption in an industry setting.

Future work will include the extension of our scheduling framework to disaggregated hardware architectures.

%% file: sections/AppendixModel.tex

This appendix presents in detail the mathematical formulation of the problem introduced in Section~\ref{sec:model}. 
In the considered setup, the CSP runs a data center that is assumed to be composed by a family $\mathcal{N}$ of nodes. 
Each node is characterized by a set $\mathcal{G}_n$ of identical GPUs. 

GPU-accelerated jobs are submitted continuously by several users; they are profiled and a suitable number of GPUs is assigned to them within a certain node. 
Each node $n \in \mathcal{N}$ can host multiple virtualized workloads, assigning in a dedicated fashion the available GPUs to jobs; a job, in turn, runs on a single node. In addition, since the number of available resources is limited, jobs that cannot be
accommodated in the system are entered into a waiting set until the next rescheduling point.  
Each node is characterized by an energy cost per unit of time $c_{ng}$ that depends on the number of used GPUs. A GPU is considered used if assigned to at least one job.
Let $\mathcal{J}$ represent the set of submitted jobs. Each job $j\in \mathcal{J}$ is characterized by a due date $d_j$, a tardiness penalty $\omega_j$, and a penalty to pay for postponing a job $\rho$.   
Furthermore, each job features an execution time $t_{jng}$ that depends on the job, the number of GPUs $g$ and their type, which in turn depends on the considered node  $n$.
The considered parameters and variables are reported in \autoref{tab:MS_PBdescrExt} for convenience.

\input{sections/MStable}

As for the variables, for node $n \in \mathcal{N}$, a variable $w_n$ determines if the node $n$ is selected, while, 
for each node $n \in \mathcal{N}$ and a number $g \in \mathcal{G}_n$ of GPUs, a variable $y_{gn}$ checks if exactly $g$ GPUs are 
used on $n$. 
The deployment of job $j$ on a specific node and number of GPUs is described, for each $j \in \mathcal{J}$ and $n \in \mathcal{N}$, by two variables: $z_{jn}$, that is equal to 1 if job $j$ is executed on node $n$, 
and $x_{jng}$, that is equal to 1 if to job $j$ are assigned, on node $n$, exactly $g$ GPUs. 
The variables representing the job tardiness and worst-case tardiness are denoted by $\tau_j$ and $\hat{\tau}_j$, respectively. The worst-case tardiness $\hat{\tau}_j$ is defined as the tardiness a job may incur in if it is postponed to the next period, that is after at most $H$ time units, using the slowest possible configuration ($M_j$).

The system execution costs are bonded to the execution costs of the first-ending jobs, thus two variables are introduced, namely $\alpha_{jn}$ and $\pi_{jn}$: the former states if job $j$ is the fastest job on node $n$ whereas the latter represents its execution cost.

The objective function of the proposed model is as follows:

\begin{footnotesize}
\[
\min \left[ \sum_{j \in \mathcal{J}}\left(\omega_j\tau_j + \rho\omega_j\hat{\tau_j}\right) + 
\sum_{\substack{j \in \mathcal{J}, n \in \mathcal{N}}}\alpha_{jn}\pi_{jn} \right].
\]
\end{footnotesize}

It aims to minimise the tardiness and the worst-case tardiness cost,
which occurs if the job is 
postponed. The second term, in turn, represents the deployment cost of the 
first-ending jobs.  Indeed, since a rescheduling can be performed every time a job completes its execution or a new job is submitted, it is reasonable to compute the objective function by considering the deployment cost of the first-ending jobs.

The resulting mathematical programming formulation is as follows:

\begin{footnotesize}
\begin{equation}\label{eq:OF}
\min \left[ \sum_{j \in \mathcal{J}}\left(\omega_j\tau_j + \rho\omega_j\hat{\tau_j}\right) + 
\sum_{\substack{j \in \mathcal{J}\\  n \in \mathcal{N}}}\alpha_{jn}\pi_{jn} \right]
\end{equation}
\end{footnotesize}

\subsubsection*{subject to}								

\begin{subequations}\label{eq:constraints2}
\scriptsize
	\begin{align}
	\label{MSha}
	&\sum_{g \in \mathcal{G}_n}x_{jng} ~\leq~ z_{jn}                    &\forall &j \in \mathcal{J}, 
	\forall n \in \mathcal{N}\\
	\label{MShc}
	&z_{jn} ~\leq~ w_n                                 &\forall &j \in \mathcal{J}, 
	\forall n \in \mathcal{N}\\
	\label{MShd}
	&x_{jng} ~\leq~ z_{jn}                              &\forall &j \in \mathcal{J}, 
	\forall n \in \mathcal{N}\\
	\label{MShe}
	&\sum_{n \in \mathcal{N}}z_{jn} ~\leq~ 1				&\forall &j \in \mathcal{J}\\
	\label{MShg}
	&\sum_{\substack{n \in \mathcal{N} \\ g \in \mathcal{G}}}x_{jng} 
	~=~ 
	\sum_{n \in \mathcal{N}}z_{jn}  							&\forall &j \in \mathcal{J}\\		
	\label{MShh}
	&\sum_{g \in \mathcal{G}_n}y_{gn} ~=~ w_n                       &\forall & n \in \mathcal{N}\\
	\label{MShi}
	&\sum_{\substack{j \in \mathcal{J}\\ \bar{g}\in \mathcal{G}_n}}\bar{g}x_{jn\bar{g}} 
	~\leq~ 
	gy_{gn} + G_n(1-y_{gn})                        &\forall & g \in \mathcal{G}_n,
	\forall n \in \mathcal{N}\\
	\label{MShj}
	&gy_{gn}
	~\leq~ 
	\sum_{\substack{j \in \mathcal{J}\\ \bar{g} \in \mathcal{G}_n}}\bar{g}x_{jn\bar{g}}                                          &\forall &g \in \mathcal{G}_n,
	\forall n \in \mathcal{N}\\
	\label{MShk}
	&\sum_{\substack{g \in \mathcal{G}_n\\ n \in \mathcal{N}}}t_{jgn}x_{jng}
	~\leq~
	d_j + \tau_j                                    &\forall &j \in \mathcal{J}\\
	\label{MShl}
	&\left(H + M_j\right)
	(1 - \sum_{n \in \mathcal{N}}z_{jn})
	~\leq~
	d_j + \hat{\tau}_j                              &\forall &j \in \mathcal{J}\\
	\label{MShm}
	&\sum_{g \in \mathcal{G}_n}t_{jgn}c_{ng}x_{jng}  
	~\leq~ 
	\pi_{jn}     													&\forall &j \in \mathcal{J},
	\forall n \in \mathcal{N}\\
	\label{MShn}
	&\sum_{j \in \mathcal{J}}\alpha_{jn} ~=~ w_n                   &\forall &n \in \mathcal{N}\\
	\label{MSho}
	&\alpha_{jn} ~\leq~ z_{jn}                           &\forall &j \in \mathcal{J},
	\forall n \in \mathcal{N}\\
	\label{MShp}
	&\sum_{n \in \mathcal{N}}w_n = \min\left\{N,J\right\} & &\\
	\label{MShq}
	&w_n \in \{0,1\}                                     &\forall &n \in \mathcal{N}\\
	\label{MShr}
	&y_{gn} \in \{0,1\}                                  &\forall &g \in \mathcal{G}_n
	\forall n \in \mathcal{N}\\
	\label{MShs}
	&z_{jn} \in \{0,1\}                                  &\forall &j \in \mathcal{J},
    \forall n \in \mathcal{N}\\
	\label{MSht}
	&x_{jng} \in \{0,1\}                                 &\forall &j \in \mathcal{J}, 
	\forall g \in \mathcal{G}_n,
	\forall n \in \mathcal{N}\\
	\label{MShu}
	&\tau_j \geq 0                                       &\forall &j \in \mathcal{J}\\
	\label{MShv}
	&\hat{\tau}_j \geq 0                                 &\forall &j \in \mathcal{J}\\
	\label{MShw}
	&\pi_{jn} \geq 0                                     &\forall &j \in \mathcal{J},
	\forall n \in \mathcal{N}\\
	\label{MShx}
	&\alpha_{jn} \in \{0,1\}                             &\forall &j \in \mathcal{J},
	\forall n \in \mathcal{N}
	\end{align}
\end{subequations}

The proposed mathematical programming formulation is non-linear, 
due to the last term of the objective function \eqref{eq:OF}, representing the 
sum of deployment costs of the first-ending jobs on each node.

 Constraints \eqref{MSha} enforce that, for all jobs $j \in \mathcal{J}$ 
and for all nodes $n \in \mathcal{N}$, at most one configuration (expressed 
by a number $g \in \mathcal{G}_n$ of GPUs) is selected and, in particular, that 
this configuration can be selected only if job $j$ is actually deployed on node 
$n$, i.e., if $z_{jn}$ is equal to 1. 
Indeed, if this happens, the sum at left-hand-side, must be less than or equal to one, meaning that at most one value $g$ can be selected as the number of GPUs assigned to job $j$ on 
node $n$.

Constraints \eqref{MShc} prescribe that a job $j$ can be deployed on 
a node $n$ only if the node is chosen. Constraints 
\eqref{MShd} enforce, in turn, that, if a job $j$ is executed on node $n$ (and 
therefore $z_{jn}$ is equal to 1), at most one number $g$ of GPUs can be 
assigned to that job, while no GPUs can be assigned to $j$ on $n$ if the job is 
not deployed on that node.

 Constraints \eqref{MShe} state that every job $j \in \mathcal{J}$ can 
be deployed at most on a node $n \in \mathcal{N}$. 

 Since $x_{jng}$ is equal to 1 only if job $j$ is executed on $n$ with 
$g$ GPUs and since, by the previous constraints, each job can be deployed at 
most on a pair node/number of GPUs, the sum at left-hand-side of Constraints \eqref{MShg} is equal to 1 if job $j$ is executed. 
The same holds for the sum at right-hand-side, since $z_{jn}$ is 1 if job $j$ is assigned to node $n$. 
The equalities, therefore, enforce that the two variables $x_{jng}$ and $z_{jn}$ give 
the same information about the execution of job $j$.

Constraints \eqref{MShh} prescribe that exactly one 
configuration, represented by a number $g \in \mathcal{G}_n$ of GPUs, must be 
selected on each node $n \in \mathcal{N}$ if the corresponding node is chosen.

 The sum at the left-hand-side of Constraints \eqref{MShi} corresponds 
to the total number of GPUs assigned to all jobs on a given node. Indeed, 
variable $x_{jn\bar{g}}$ is equal to 1, for a fixed job $j$ and node $n$, only 
if that job is deployed on that node with exactly $\bar{g}$ GPUs. Since, 
for each running job, only one number $\bar{g}$ of GPUs can be selected, 
as prescribed by the previous constraints, the sum over $\bar{g} \in 
\mathcal{G}_n$ of $\bar{g}x_{jn\bar{g}}$ turns out to be equal to the number of GPUs assigned 
to job $j$. The overall sum represents, therefore, the total number of GPUs 
assigned to all jobs running on $n$. This number must be less than or equal to 
$g$, if $g$ is the number of GPUs selected on node $n$, which happens 
when variable $y_{gn}$ is equal to 1. If, in turn, $y_{gn}$ is 0, the inequality prescribes that the total number of GPUs assigned to 
jobs on node $n$ is less than or equal to the maximum number 
of available GPUs on the node itself. 

On the other hand, Constraints \eqref{MShj} enforce the total number 
of GPUs assigned to jobs on node $n$ to be greater than or equal to $g$. 
Therefore, by combining \eqref{MShi} and \eqref{MShj}, the total number of 
GPUs assigned to jobs on node $n$ must be exactly equal to $g$, if $g$ is the 
number of GPUs selected on that node.

Constraints \eqref{MShk} and \eqref{MShl} are used to define the 
tardiness and the worst-case tardiness of every job $j \in \mathcal{J}$. In 
particular, Constraints \eqref{MShk} state that the total execution time of 
job $j$ on the selected configuration must be less than or equal to the sum 
between its due date and its tardiness, which entails that $\tau_j$ is 
greater than or equal to the execution time of job $j$ minus its due date. 
At the left-hand-side of Constraints \eqref{MShl}, the sum over all 
nodes of $z_{jn}$ is equal to 1 when job $j$ is executed and it is 0 when job $j$ is postponed, since $z_{jn}$ is 0 for all nodes. In the first 
case, the sum between the due date and the worst-case tardiness of job $j$ is 
simply prescribed to be a non-negative number. In the second case, 
the worst-case tardiness is defined as the sum of the scheduling interval 
$H$ and the maximum execution time of job $j$, minus its due date.

Constraints \eqref{MShm} prescribe the deployment cost of job $j$ on 
node $n$ to be greater than or equal to the energy cost of GPUs used on that 
node, multiplied by the amount of time spent by job $j$ on $n$. Since 
$x_{jng}$ is equal to 1 if job $j$ is executed on $n$ with $g$ GPUs and since 
only one of those variables can be equal to one for each job, the sum 
over all $g \in \mathcal{G}_n$ of $t_{jgn}x_{jng}$ is equal to the execution time 
of $j$ with the unique number of GPUs it is assigned to. Moreover, the term $t_{jgn}x_{jng}$ is multiplied by $c_{ng}$ to compute the energy cost of the selected configuration.

Constraints \eqref{MShn} and \eqref{MSho} state that there must be, 
on each selected node, exactly one $j \in \mathcal{J}$ being the 
first-ending job. Constraint \eqref{MShp}, where $J$ and $N$ denote 
the cardinalities of $\mathcal{J}$ and $\mathcal{N}$, respectively, is 
used to enforce the usage of nodes. Jobs that are postponed are indeed more likely to 
violate their due date. In order to avoid the resulting penalty costs, and since 
job deployments can be redefined in a following rescheduling thanks to 
preemption, it is always better to execute jobs as soon as resources are 
available. Finally, Constraints \eqref{MShq} to \eqref{MShx} define the 
domain of the variables.

%% file: sections/MStable.tex

\begin{table}[t]
	\caption{Problem parameters and variables}
	\footnotesize
	\label{tab:MS_PBdescrExt}
	\begin{tabular*}{\columnwidth}{| l | l |} 
		\hline
		\multicolumn{2}{| l |}{\textbf{Parameters}}\\   
		\hline
		$\mathcal{J}$          & set of submitted jobs \\
		$\mathcal{N}$          & set of available nodes \\
		$\mathcal{G}_n$           & set of available GPUs in each node $n \in \mathcal{N}$\\
		$G_n$           			& number of available GPUs in each node $n \in \mathcal{N}$\\
		$c_{ng}$       & energy cost of each node $n \in \mathcal{N}$ when $g$ GPUs are used \\
		$r_j$		& revenue for executing job $j \in \mathcal{J}$ \\
		$d_j$       & deadline of job $j \in \mathcal{J}$ \\
		$\omega_j$  & tardiness weight of job $j \in \mathcal{J}$ \\
		$M_j$       & maximum execution time of job $j \in \mathcal{J}$ \\
		$t_{jng}$    & execution time of job $j \in \mathcal{J}$ when running on $g$ GPUs on \\
		 & node $n \in \mathcal{N}$\\
		$\hat{M}_j$ & maximum possible deployment cost for job $j \in \mathcal{J}$ \\
		$\rho$             & penalty coefficient for postponed jobs\\
		$H$           & scheduling time interval \\
		\hline
		\multicolumn{2}{| l |}{\textbf{Variables}}\\    
		\hline
		$w_n$       & 1 if node $n \in \mathcal{N}$ is chosen \\
		$y_{gn}$    & 1 if $g$ GPUs are used on node $n \in \mathcal{N}$ \\
		$z_{jn}$    & 1 if job $j \in \mathcal{J}$ is executed on node $n \in \mathcal{N}$ \\
		$x_{jng}$   & 1 if job $j \in \mathcal{J}$ is executed on node $n \in \mathcal{N}$ with
		$g$ GPUs \\
		$\tau_j$    & tardiness of job $j \in \mathcal{J}$ \\
		$\hat{\tau}_j$    & worst-case tardiness of job $j \in \mathcal{J}$ \\
		$\pi_{jn}$  & operation cost of job $j \in \mathcal{J}$ on node $n \in \mathcal{N}$ \\
		$\alpha_{jn}$  & 1 if job $j \in \mathcal{J}$ is the first ending job on node 
		$n \in \mathcal{N}$ \\
		\hline
	\end{tabular*}
\end{table}

%% file: sections/AppendixExperiments.tex
All experiments described in this paper  have been performed relying on ARMIDA (ARM Infrastructure for the Development of Applications) infrastructure, a test \& development 8-node cluster located at E4’s premises. Each node has 2xMarvell TX2 sockets (per socket: 64 cores Marvell TX2@2,2/2.5 GHz, 256 GB RAM, 1xMellanox IB 100Gb EDR). ARMIDA is based on RedHat Enterprise Linux 8.0 (kernel 4.18), GNU 8.3.1. The installed software includes ARM Allinea Studio (Compiler + Forge), GCC 10, Nvidia HPC SDK, CUDA 11. Specifically, the characteristics of the ARMIDA nodes in terms of available types and numbers of GPUs are summarized in \autoref{tab:catalog}.

To validate the results provided by ANDREAS Optimizer in the ARMIDA setup, we have considered a prototype system including 8 different jobs. Their characteristics are summarized in \autoref{tab:jobs},
where we denote by $S_j$, $d_j$ and $\omega_j$ the submission time, due date and tardiness weight of each job, respectively.

\input{tables_and_algorithms/jobs}

\begin{figure}[ht]
	\centering
	\includegraphics[width=\columnwidth]{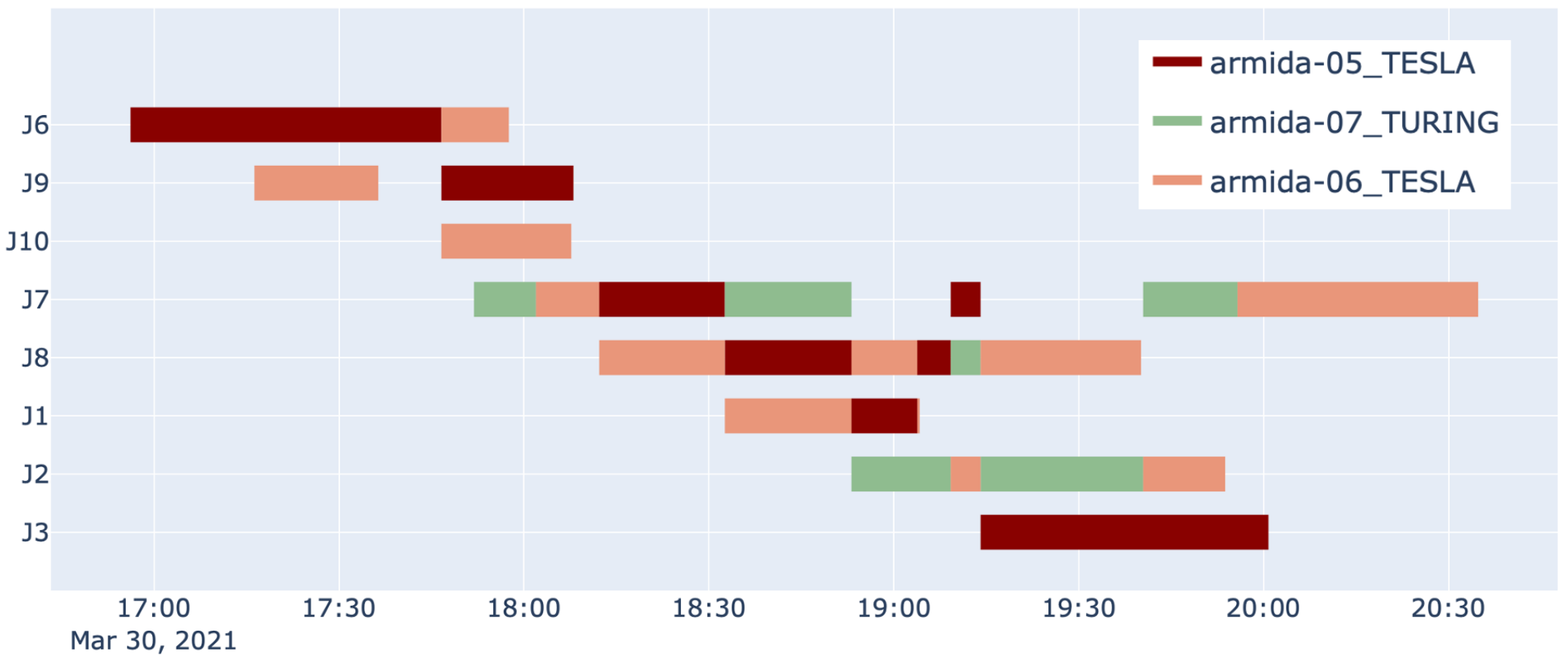}
	\caption{Trace of jobs execution}
	\label{fig:RealTrace}
\end{figure}

\autoref{fig:RealTrace} reports the sequence of jobs executions, along with the relative configurations. Jobs are submitted with a fixed inter-arrival time of $1200s$, and the Job Manager invokes the Job Optimizer periodically every 5 minutes. This time has been selected to avoid too frequent reoptimizations, which may dictate the necessity of migrating jobs among different GPUs, thus incurring in additional costs, but also to reduce jobs waiting times and to prevent the machines from remaining idle for too long when the resources are released. From the trace, it is possible to observe how the two GPUs available on node armida-06 are shared, when possible, among different jobs (as it happens, e.g., with J6 and J10 when the latter is submitted). Conversely, jobs normally receive both the GPUs, which guarantees a faster execution, when their due date is approaching. Moreover, it is possible to observe the preemption mechanism, that comes into place, for instance, when J2 is submitted: since it has a stricter due date than J7, it is immediately deployed on armida-07, while the execution of the latter is postponed to a future scheduling point.

\begin{figure}[ht]
    \centering
	\includegraphics[width=\columnwidth]{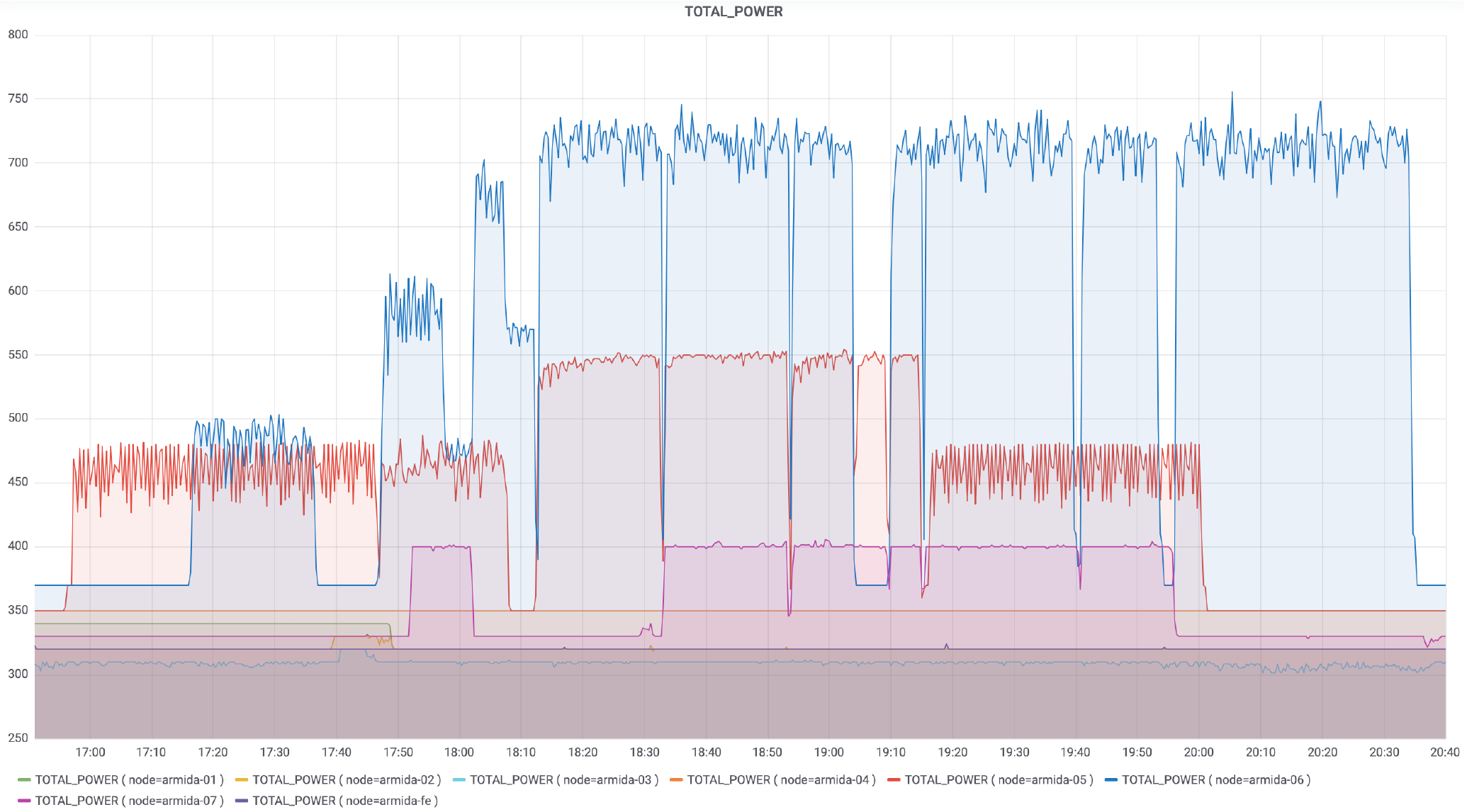}
	\caption{Power consumption on ARMIDA nodes}
	\label{fig:RealCosts}
\end{figure}

The power consumption of ARMIDA nodes during the full experiment session can be observed in \autoref{fig:RealCosts}. This has been used to compute the actual energy costs incurred by the system during the whole experiment, including the reconfiguration costs that are not considered by our simulator.

%% file: tables_and_algorithms/jobs.tex

\begin{table}[ht]
    \centering
    \begin{scriptsize}
    \caption{Applications submitted to the system}
    \label{tab:jobs}
    \centering
    \begin{tabular}{|l|l|r|r|r|r|r|}
        \hline
        \textbf{Job}    & \textbf{Application}   &  \textbf{Epochs}   &  \textbf{Batch}   &  \textbf{$S_j$ [s]}   &  \textbf{$d_j$ [s]}   &  \textbf{$\omega_j$}\\
        \textbf{ID}    & \textbf{}   &  \textbf{}   &  \textbf{size}   &  \textbf{}   &  \textbf{}   &  \textbf{}\\
        \hline
        J6   & EfficientNet & 80    & 4096  & 0 & 3600   & 4\\
        J9   & ConvNet & 160    & 8192  & 1200 & 2600   & 2\\
        J10   & ConvNet & 80    & 8192  & 2400 & 7600   & 3\\
        J7   & LSTM-big & 160    & 8192  & 3600 & 17600   & 3\\
        J8   & LSTM-small & 160    & 8192  & 4800 & 7600   & 3\\
        J1   & LSTM-big & 60    & 8192  & 6000 & 5600   & 5\\
        J2   & LSTM-small & 60    & 8192  & 7200 & 12600   & 2\\
        J3   & EfficientNet & 60    & 4096  & 8400 & 11600   & 1\\
        \hline
    \end{tabular}
    \end{scriptsize}
\end{table}